# Induced-charge electro-osmosis around metal and Janus spheres in water: Patterns of flow and breaking symmetries

Chenhui Peng, Israel Lazo, Sergij V. Shiyanovskii and Oleg D. Lavrentovich

Liquid Crystal Institute and Chemical Physics Interdisciplinary Program,
Kent State University, Kent, OH 44242

**Abstract**

We establish experimentally the flow patterns of induced-charge electro-osmosis (ICEO) around immobilized metallic spheres in aqueous electrolyte. The AC field modifies local electrolyte concentration and causes quadrupolar flows with inward velocities being smaller than the outward ones. At high fields, the flow becomes irregular, with vortices smaller than the size of the sphere. Janus metallo-dielectric spheres create dipolar flows and pump the fluid from the dielectric toward the metallic part. The experimentally determined far-field flows decay with the distance as $r^{-3}$.

Electrically driven flows of fluids and transport of particles, collectively called electrokinetics [1], is a fascinating area of research receiving a renewed attention thanks to the advances in microfluidics, studies of charge transport in organic matter, and science of active matter [2-4]. Of special interest is induced–charge elecroosmosis (ICEO), in which an applied field $E$ generates a charge near polarizable (metal) surfaces and then acts on this charge to drive a fluid flow [3]. The amplitude of ICEO velocities is on the order of [5]

$$u_0 = \frac{\varepsilon \varepsilon_0 a}{\eta} E^2 \tag{1}$$

where $\varepsilon$ and $\eta$ are permittivity and viscosity of the medium, respectively, $\varepsilon_0$ is the electric constant, $a$ is the particle's radius. One power of $E$ induces the charge, while the second drives the resultant ICEO flow [5], which explains the quadratic dependence $u_0 \propto E^2$. The product $aE$ in Eq.(1) can be considered as an effective zeta potential created by the electric field. ICEO is thus dramatically different from the classic linear electro-osmosis in which the zeta potential, determined by fixed surface charges, does not depend on $E$, so that $u_0 \propto E$. The quadratic dependence $u_0 \propto E^2$ allows one to drive steady flows with a uniform AC field, a clear advantage over DC driving of the standard linear electro-osmosis, which explains the growing interest to ICEO.

Nonlinear electro-osmosis around conductive bodies in a uniform field have been documented experimentally, for mercury drops [6], metal spheres [7], metallized "volcano" posts [8] and metallic cylinders [9-10]. In all these cases, the flows are quadrupolar and produce no net pumping of the fluid. Bazant and Squires made an important prediction that ICEO combined with broken symmetry of particles can result in an AC-driven pumping and electrophoresis of free particles [11-12]. Discovery of AC-electrophoresis of Janus particles [13] and spinning of Janus doublets [14] supports the idea that broken symmetry of the particle can lead to ICEO pumping and electrophoresis. Pumping has been also observed in a somewhat different context of asymmetric electrodes subject to AC voltage [15-16].

Current understanding of ICEO and role of symmetry breaking is far from being complete, however. Even for the simple case of a homogeneous metallic sphere, it is unclear why the experimental velocities [10,13] are 10-100 times smaller than in Eq.(1), especially when voltages applied across the particle, $\propto Ea$, exceed the thermal voltage $V_{th} = k_B T / e = 25$ mV. Recent numerical simulations [17] suggest that one of the reasons might be the concentration polarization effect [18-19]. Besides the somewhat reduced velocities, the simulations [17] reveal other features not anticipated by the standard theory and not seen in prior experiments, such as asymmetry of inward and outward velocities and a "chaotic" behavior at high voltages $\geq 30 V_{th}$.

Comparison of the available theoretical, numerical and experimental works makes it clear that what is lacking in the field of AC-driven electrokinetics is a direct experimental determination of how the ICEO velocity and ionic concentration depend on spatial coordinates around a polarizable sphere and how these patterns change with symmetry of the sphere's properties. We present such an experiment for (a) a conductive sphere and (b) a metal-dielectric Janus sphere, both placed in a flat microfluidic chamber with an aqueous electrolyte.

The conductive spheres (a) were obtained by depositing a ~200 nm layer of gold (Au) onto dry soda lime glass spheres of diameter $2a = (50 \pm 2)$ μm (*ThermoScientific*). Janus spheres (b) were created by covering only half of the glass spheres with Au. Microfluidic chambers of thickness $h = 50; 60; 170;$ and 500 μm were formed by two parallel glass plates, kept apart by parallel aluminum electrodes (separation $L = 10$ mm). The spheres were immobilized at the bottom plate by *Norland* adhesive and separated by large distances (300-800 μm) from each other. Most experiments were performed with deionized ultrapure water of conductivity $\sigma \approx 10^{-5}$ S/m, since it has been reported that pure water yields the highest electrophoretic velocity of Janus particles while addition of salts cause a slow-down [13]. We also used water with added 1 mM KCl; the difference in measured velocities was less than 5%. The AC (sinusoidal) field was applied in the plane of the cell, with the typical amplitude $E \sim 10^4$ V/m. Small conductivities and applied fields allow us to avoid the effects of Joule heating: As reported by Ramos et al. [20], temperature increase of water electrolytes of conductivity $\sigma \approx 10^{-2}$ S/m caused by fields $E \approx 10^6$ V/m is only about 0.2°C in the broad frequency range between 1 kHz and 20 MHz. As the function of field frequency $f$, the ICEO velocities around Au spheres reach a maximum near 1 kHz.

The ICEO velocities are measured under the microscope Nikon Eclipse E600 equipped with CARV confocal imager (BD Biosciences, Inc.) and video camera Photometrics Cascade 650, using micro-particle imaging velocimetry ($\mu$PIV) [21]. The tracers are fluorescent polystyrene spheres (*Bangs Laboratories*) added in small quantities ($\sim 0.01$ wt%). The tracers (diameter $2R = 1\mu$m) are much smaller than the Au spheres, in order to avoid dielectrophoretic effects [9]. Under the DC driving, the tracers show linear electrophoresis with a velocity $\sim 10\,\mu$m/s at $E = 0.5$mV/$\mu$m. At 1 kHz AC driving, electrophoretic shifts of the tracers are submicron, i.e. much smaller than the displacements caused by ICEO. For $E \leq 12$ mV/$\mu$m, the fluorescent signal was recorded with an exposure time $\Delta \tau = 40$ ms; the velocity patterns were obtained by superimposition of over 1500 images using a $\mu$PIV software [21]. For higher fields, we used $\Delta \tau = 20$ ms and obtained instantaneous velocities by superimposing pairs of images. In a parallel experiment, to map the concentration of ions, we used charged fluorescent dye

Rhodamine 6G added in small quantities (0.1 mM). The intensity of fluorescence was linearly proportional to the dye concentration $c$ in the range 0.01-0.14 mM. The microscope was focused at the equatorial plane of the spheres.

*Homogeneous Au spheres.* The experimental ICEO flows and velocity maps, Fig.1, show quadrupolar symmetry with four large steady vortices. Although the pattern is similar to the prediction of the standard ICEO theory, there are important qualitative and quantitative differences.

The standard ICEO theory predicts that the radial component of velocity depends on the polar angle $\theta$ measured with respect to the field $\mathbf{E}=(E,0,0)$: $u_r \propto 1+3\cos 2\theta$ when the conductive particle is a sphere and $u_r \propto \cos 2\theta$ when it is a cylinder [5]. The inward $u_{r,in} = u_r(\theta = 0, \pi)$ velocity is expected to be either larger than the outward velocity $u_{r,out} = u_r(\theta = \pm \pi/2)$, or equal to it, namely, $R_{io}^u = |u_{r,in}|/|u_{r,out}| = 2$ for the sphere and $R_{io}^u = 1$ for the cylinder. For spheres in confined cells such as ours, the expected $R_{io}^u$ is between 1 and 2. The experimental $R_{io}^u$ turns out to be noticeably smaller, Fig.1d,g.

The inward/outward asymmetry is readily seen in the Fourier analysis of the experimental patterns, Fig.1a,b. The radial $u_r(r,\theta)$ and azimuthal $u_\theta(r,\theta)$ velocity components can be represented in polar coordinates $(r,\theta)$ as

$$u_r(r,\theta) = \sum_{n=1}^{\infty} u_r^{(n)}(r)\cos n\theta, \quad u_\theta(r,\theta) = \sum_{n=1}^{\infty} u_\theta^{(n)}(r)\sin n\theta, \qquad (2)$$

where $r$ is the distance from the sphere's center. The experimental velocities are well captured by Eqs. (2) with two even harmonics $n = 2, 4$: $u_r(r,\theta) = u_r^{(2)}(r)\cos 2\theta + u_r^{(4)}(r)\cos 4\theta$ and $u_\theta(r,\theta) = u_\theta^{(2)}(r)\sin 2\theta + u_\theta^{(4)}(r)\sin 4\theta$, Fig.1c,d,e. Other harmonics are negligibly small. The coefficients $u_r^{(2)}$ and $u_r^{(4)}$ are of opposite signs, Fig.1d, which makes the ratio $R_{io}^u = \left|u_r^{(2)} + u_r^{(4)}\right| / \left|-u_r^{(2)} + u_r^{(4)}\right|$ *smaller* than 1. The data in Fig.1d taken for $h = 60\,\mu\text{m}$, yield $R_{io}^u \approx 0.58$ at $r = 35\,\mu\text{m}$ and $\approx 0.67$ at $r = 43\,\mu\text{m}$. For smaller $h = 2a = 50\,\mu\text{m}$, the effect is even stronger, as $R_{io}^u \approx 0.5$ at $r = 35\,\mu\text{m}$ and $\approx 0.62$ at $r = 43\,\mu\text{m}$. For thicker samples, $h = 170\,\mu\text{m}$, we find $R_{io}^u \approx 0.7$ at $r = 35\,\mu\text{m}$ and $\approx 0.76$ at $r = 43\,\mu\text{m}$. Finally, for $h = 500\,\mu\text{m}$, $R_{io}^u \approx 0.94$ at $r = 35\,\mu\text{m}$ and $\approx 1.0$ at $r = 43\,\mu\text{m}$. Small values of $R_{io}^u$ are observed in different cells and are not associated with imperfections such as nonuniform thickness of the cell.

The four Fourier coefficients show non-monotonous radial dependence, Fig.1d,e. At large distances $r \gg a$, they decay as $u_{r,\theta}^{(n)}(r) \propto r^{-3}$, Fig. 1e, and obey the condition of incompressibility $\nabla \cdot u\big|_{r \gg a} = 0$, i.e., $u_r^{(2)} = u_\theta^{(2)}$ and $u_r^{(4)} = 2u_\theta^{(4)}$. The same behavior is observed for thinner and thicker cells.

The higher is the field, the stronger is the asymmetry of the inward and outward flows. The ratio of inward to outward maximum velocities plotted in Fig.1g decreases from 0.8 at $E = 6\,\text{mV}/\mu\text{m}$ to 0.4 at $E = 19\,\text{mV}/\mu\text{m}$, suggesting again that the observed asymmetry is not associated with cell imperfections and can be caused by concentration polarization. Our experiments with charged Rhodamine 6G added at $c = 0.1$ mM indeed show that the concentration of ions, while spatially uniform without a field, is re-distributed when the field is present, increasing by about 7% near the sphere at $\theta = 0, \pi$ and decreasing at $\theta = \pm\pi/2$, Fig.2.

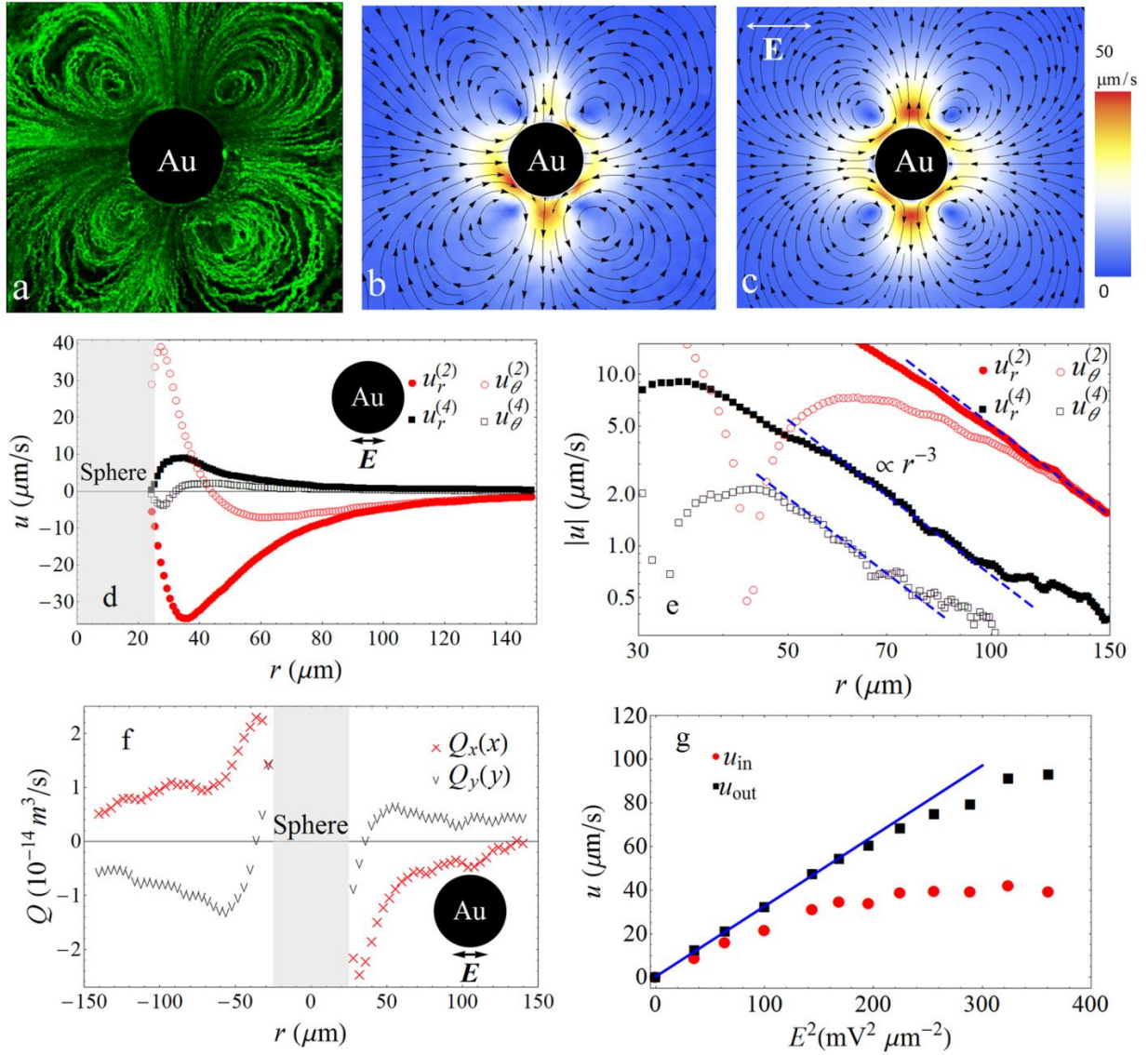

**Figure 1.** (Color online) (a-f) AC electric field ($E = 10\,\text{mV}/\mu\text{m}$, $f = 1\,\text{kHz}$) induced ICEO around Au sphere ($2a = 50\,\mu\text{m}$) in $h = 60\,\mu\text{m}$ chamber filled with water. (a) ICEO trajectories. (b) Experimental velocities map. (c) Numerically reconstructed velocity map with Fourier harmonics $n=2$ and 4, Eq.(2). (d,e) Radial dependence of Fourier coefficients in linear (d) and log-log (e) presentation. (f) Volumetric flow through the cross-sections of the cell. (g) Maximum inward $|u_{in}|$ and outward $|u_{out}|$ velocities as a function of $E^2$. Dashed lines in (e) show a $1/r^3$ slope.

At $E = 10\,\text{mV}/\mu\text{m}$, $h = 60\,\mu\text{m}$, the maximum ICEO velocity $|u^{\max}| = |u_{out}|$ reaches $50\,\mu\text{m/s}$ at $r \approx 30\text{-}40\,\mu\text{m}$. For low voltages, $|u_{out}|, |u_{in}| \propto E^2$, but above ~8 mV/μm, the velocities grow slower than $E^2$, Fig. 1g. Beginning with ~20 mV/μm, the flows become unsteady, with vortices of a small size, ~$(0.1-0.3)a$, nucleating and moving in a chaotic manner near the sphere, Fig.3.

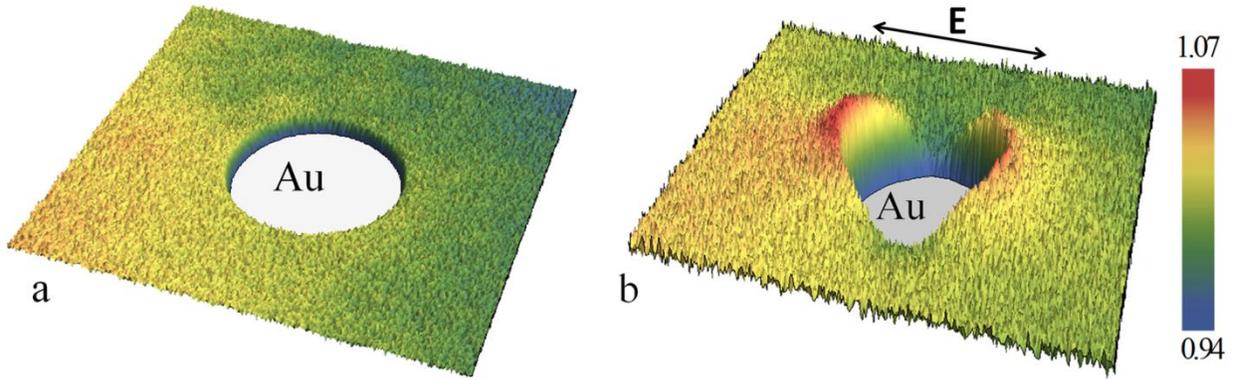

**Figure 2.** (Color online) Experimental map of fluorescence intensity caused by positively charged dye Rhodamine 6G at (a) zero electric field; (b) electric field $E = 15\,\text{mV}/\mu\text{m}$, 1 kHz.

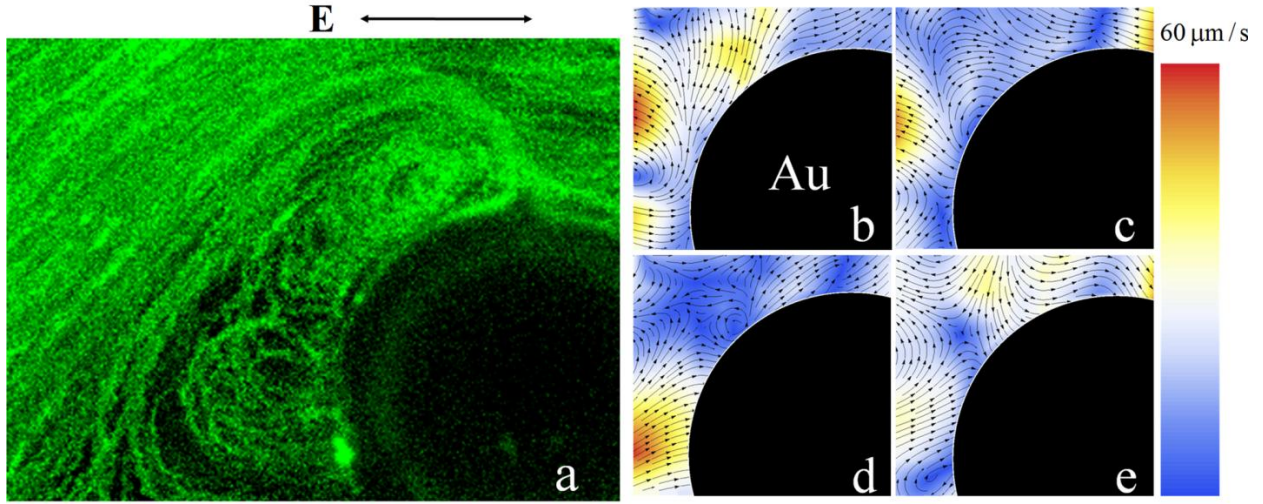

**Figure 3.** (Color online) Experimental ICEO (a) trajectories and (b,c,d,e) velocity patterns measured over randomly selected 40 ms time intervals, around a homogeneous Au sphere. $E = 40\,\text{mV}/\mu\text{m}$, $f = 8$ kHz, $2a = 50\,\mu\text{m}$.

If a glass sphere is used instead of the Au sphere, the ICEO velocities are much smaller, ~ $0.1\,\mu m/s$ even for $E = 80\,mV/\mu m$. This result is natural, as the velocity around a dielectric particle is determined by the Debye screening length $\lambda_D$ [5] that is much smaller than $a$.

*Janus Au-SiO$_2$ spheres.* We introduce a structural dipole, directed from the SiO$_2$ part towards the Au part. The geometry $\mathbf{E} \perp \mathbf{P}$ corresponds to the orientation of a free Janus particle in an external field [13]. ICEO around an immobilized Janus sphere breaks left-right symmetry, creating a pair of strong vortices near the Au hemisphere, Fig. 4a,b. The velocities near the Au part are much higher than near the SiO$_2$ part, because of higher polarizability of Au [12]. Only two harmonics, $n = 1, 2$, are needed to reproduce the experimental Fig. 4c. At large distances, the second harmonic exhibits a cubic decay $u_{r,\theta}^{(2)}(r) \propto r^{-3}$ and satisfies the condition of incompressibility, $u_r^{(2)}(r) = u_\theta^{(2)}(r)$. The first harmonic shows a slower decay.

Presence of the $n = 1$ harmonic implies pumping of water around an immobilized Janus particle (and explains AC electrophoresis of a free particle observed in [13]). To quantify the efficiency of pumping, we define the volumetric flow passing through cross-sections of the cell, Fig.4e, as a function of their distances from the sphere, $Q_x(x) = \frac{2}{3} h \int_{-y_0}^{y_0} u_x(x,y) dy$ for the $yz$ sections and $Q_y(y) = \frac{2}{3} h \int_{-x_0}^{x_0} u_y(x,y) dx$ for the $xz$ sections. The coefficient 2/3 reflects the parabolic velocity profile along the $z$ axis. Here $x_0 = y_0 = 150\,\mu m$; $h = 60\,\mu m$; the functions $u_x(x,y)$ and $u_y(x,y)$ are known from Fig.4b. $Q_x$ is positive for both positive and negative $x$. The same quantity $Q_x$ measured around the homogeneous Au sphere, shows an asymmetric behavior, as there is no pumping in this case, Fig.1f.

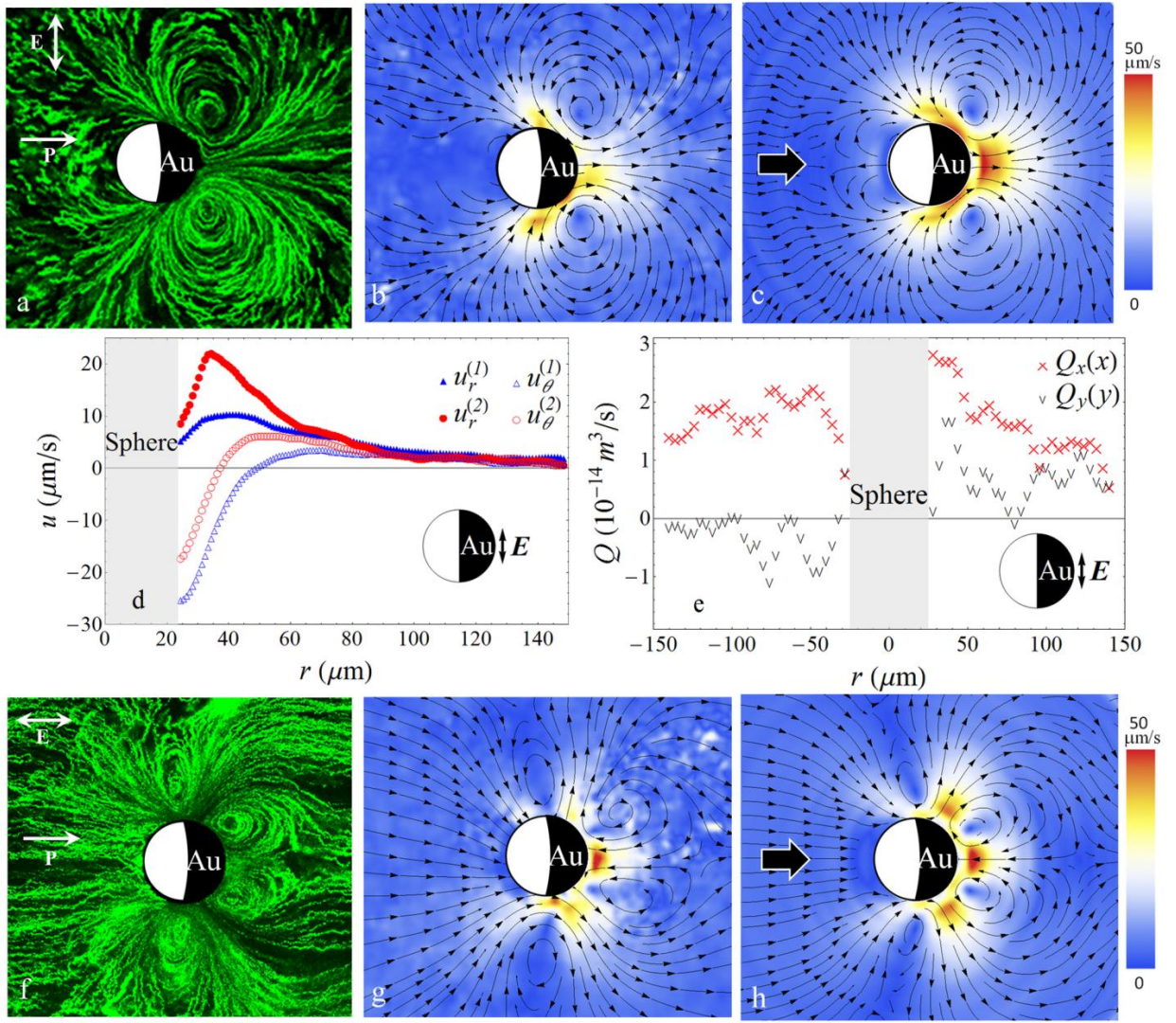

**Figure 4.** (Color online) ICEO flows around Janus particles with (a-e) $\mathbf{E} \perp \mathbf{P}$ and (f-h) $\mathbf{E} \parallel \mathbf{P}$. (a,f) ICEO trajectories; (b,g) Experimental ICEO velocities maps. (c,h) Numerically reconstructed velocity maps with two Fourier harmonics $n=1,2$ in (c) and four harmonics $n=2,3,4$ in (h). (d) Radial dependence of coefficients for the case (b,c). (e) Volumetric flows.

If $\mathbf{E} \parallel \mathbf{P}$, fore-aft symmetry is broken, Fig.4f,g,h. In weak fields, $E \approx 5\,\mathrm{mV/\mu m}$, there are two vortices near the Au hemisphere; similarly to the case of $\mathbf{E} \perp \mathbf{P}$. At $E \approx 10\,\mathrm{mV/\mu m}$, however, a second pair of vortices appears near the Au/SiO$_2$ interface, Fig. 4f,g,h. This secondary pair has been predicted by numerical studies [22] and is apparently associated with a condensed layer of free charges at the metal-dielectric boundary. For $E \geq 10\,\mathrm{mV/\mu m}$, an accurate reproduction of Fig.4g is achieved only when four harmonics are involved, $n = 1, 2, 3$, and 4, Fig.4h.

With the typical $\varepsilon = 80$, $\eta = 1 \, \text{mPa} \cdot \text{s}$, $E = 10 \, \text{mV/}\mu\text{m}$ and $a = 25 \, \mu\text{m}$, Eq. (1) predicts $u_0 \approx 2 \times 10^3 \, \mu\text{m/s}$, while the experimental $|u^{\max}| = |u_{out}| \approx 50 \, \mu\text{m/s}$ is much lower, by a factor $\Lambda = u_0 / |u^{\max}| \approx 40$. There are many potential reasons. One is concentration polarization [17], supported by our experiments in two aspects. First, in the moderate regime, $Ea = 10 V_{th}$, the inward velocities are smaller than the outward velocities, Fig.1g. Second, we do observe a field-induced redistribution of ions, Fig.2. However, the velocity reduction factor $\Lambda_{cp}$ caused by concentration polarization has been predicted to be only about 5 [17], thus the effect is unlikely to be the sole factor responsible for the smallness of velocities. Among other reasons are surface contamination [23] and roughness [24], interactions with walls [13], confinement-enhanced viscous friction [25] and the assumption of small voltages $V \leq V_{th}$ of the standard theory.

To summarize, we have experimentally described the ICEO flows around immobilized spheres and demonstrated that these patterns are controlled by broken symmetries of surface properties. Homogeneous spheres produce quadrupolar flows with a subtle effect of the inward velocities being smaller than the outward velocities. The effect is consistent with the idea of concentration polarization. Field-induced redistribution of charged fluorescent dye around homogeneous Au spheres, Fig.2, provides another evidence of concentration polarization. At high fields, an irregular ICEO pattern develops around the spheres, which might be caused by many effects, from surface roughness to concentration polarization and Faradaic reactions. Metal-dielectric Janus spheres produce dipolar flows and pumping effects, with contributions $\sim \cos\theta$, $\sim \sin\theta$ to the radial and azimuthal components of ICEO velocities, respectively.

ICEO-mediated pumping around immobilized particles and nonlinear electrophoresis of free Janus spheres [13] are both rooted in broken symmetry of the particle properties (such as shape or metal-dielectric duality of the Janus spheres) [11-12]. A complementary set of phenomena is observed in anisotropic (liquid crystalline) electrolytes, in which it is the broken symmetry of the medium (rather than of a particle) that brings about pumping, mixing [26], and nonlinear electrophoresis [27]. For example, a liquid crystal with broken symmetry of molecular orientation enables electrophoretic motion of a particle even if it is of a perfect spherical shape [27-28]. Combination of both mechanisms, with broken symmetry of particles and broken symmetry of the medium, might bring new interesting results in the rich field of charge transport in organic matter and optimize strategies of mixing, pumping, and micron-scale transport. Such a work is in progress.

We gratefully acknowledge the NSF DMR grant 1104850 for support of this work. ODL thanks S. M. Davidson, A. Mani, N. A. Mishchuk, and T. M. Squires for illuminating discussions.

**References**

[1] H. Morgan and N. G. Green, *AC electrokinetics: colloids and nanoparticles* (Research Studies Press Ltd. , Baldock, 2003), Mirotechnology and Microsystems series.
[2] M. Z. Bazant, M. S. Kilic, B. D. Storey, and A. Ajdari, Advances in Colloid and Interface Science **152**, 48 (2009).
[3] M. Z. Bazant and T. M. Squires, Current Opinion in Colloid & Interface Science **15**, 203 (2010).
[4] I. S. Aranson, Physics-Uspekhi **56**, 79 (2013).
[5] T. M. Squires and M. Z. Bazant, Journal of Fluid Mechanics **509**, 217 (2004).
[6] V. Murtsovkin and G. Mantrov, Colloid Journal of the USSR **53**, 240 (1991).
[7] N. Gamayunov, G. Mantrov, and V. Murtsovkin, Colloid journal of the Russian Academy of Sciences **54**, 20 (1992).
[8] C. K. Harnett, J. Templeton, K. A. Dunphy-Guzman, Y. M. Senousy, and M. P. Kanouff, Lab on a Chip **8**, 565 (2008).
[9] J. A. Levitan, S. Devasenathipathy, V. Studer, Y. Ben, T. Thorsen, T. M. Squires, and M. Z. Bazant, Colloids and Surfaces A: Physicochemical and Engineering Aspects **267**, 122 (2005).
[10] C. Canpolat, S. Qian, and A. Beskok, Microfluid Nanofluid **14**, 153 (2013).
[11] M. Z. Bazant and T. M. Squires, Physical Review Letters **92**, 066101 (2004).
[12] T. M. Squires and M. Z. Bazant, Journal of Fluid Mechanics **560**, 65 (2006).
[13] S. Gangwal, O. J. Cayre, M. Z. Bazant, and O. D. Velev, Physical Review Letters **100**, 058302 (2008).
[14] A. Boymelgreen, G. Yossifon, S. Park, and T. Miloh, Physical Review E **89**, 011003 (2014).
[15] A. B. D. Brown, C. G. Smith, and A. R. Rennie, Physical Review E **63**, 016305 (2000).
[16] P. García-Sánchez, A. Ramos, N. Green, and H. Morgan, IEEE Transactions on Dielectrics and Electrical Insulation **13**, 670 (2006).
[17] S. M. Davidson, M. B. Andersen, and A. Mani, Physical Review Letters **112**, 128302 (2014).
[18] S. S. Dukhin, Advances in Colloid and Interface Science **35**, 173 (1991).
[19] N. A. Mishchuk, Advances in Colloid and Interface Science **160**, 16 (2010).
[20] A. Ramos, H. Morgan, N. Green, and A. Castellanos, Journal of Physics D: Applied Physics **31**, 2338 (1998).
[21] W. Thielicke and E. J. Stamhuis, Time Resolved Digital Particle Image Velocimetry Tool for MATLAB, http://pivlab.blogspot.com/p/pivlab-documentation.html. (2010).
[22] Y. Daghighi, Y. Gao, and D. Li, Electrochimica Acta **56**, 4254 (2011).
[23] A. J. Pascall and T. M. Squires, Physical Review Letters **104**, 088301 (2010).
[24] R. J. Messinger and T. M. Squires, Physical Review Letters **105**, 144503 (2010).
[25] J. Happel and H. Brenner, *Low Reynolds number hydrodynamics: with special applications to particulate media* (Martinus Nijhoff Publishers, The Hague, 1983), Vol. 1.
[26] I. Lazo, C. Peng, J. Xiang, S. V. Shiyanovskii, and O. D. Lavrentovich, Nature Communications **5**, 5033 (2014).
[27] O. D. Lavrentovich, I. Lazo, and O. P. Pishnyak, Nature **467**, 947 (2010).
[28] S. Hernandez-Navarro, P. Tierno, J. Ignes-Mullol, and F. Sagues, Soft Matter **9**, 7999 (2013).